# Properties of Superconductive Nb and NbTi Coatings-Films Deposited by New Ionic-Plasma-Clusters Method.

K. Pavlov [a],* A. Savin [b], J. Laurila [c], P. Vuoristo [c], K. Gorchakov [a], S. Gorchakova [a].

[a] - Kraftonweg Oy, Ruuvikatu 6, 48770, Kotka, Finland
[b] - Low Temperature Laboratory, Department of Applied Physics, Aalto University, PO Box 15100, FI-00076 AALTO, Finland
[c] - Tampere University of Technology Faculty of Engineering Sciences Laboratory of Materials Science Surface Engineering Research Group P.O. Box 589, FIN-33101 Tampere, FINLAND

**Abstract.**

*Originally developed technique based on generation of plasma-cluster flux emitted from cathode spots at surface of an integrally cold cathode has been applied for deposition of niobium superconductive thick film materials. Possibility for depositing thick film coatings in homogeneous, heterogeneous and layered structures is shown to be feasible. Experimental results of mechanical tests including Tensile Testing resulting in stretching of coatings are presented. Critical parameters of deposited superconducting films ($T_c$, $J_c$ and $B_c$) with various types of structural compositions have been investigated. Niobium- titanium layered coating films deposited by ionic-plasma-clusters method are characterized by very high values of critical current density. All results reveal coating material high value in terms of manufacturability. High reliability, low production costs and ease of implementation of new technology for a wide-scale production of superconducting film materials have been demonstrated.*

**Introduction**

Nowadays the implementation of superconducting materials in modern industry is limited due to their relatively high cost and extreme complexity to process. Most widely used materials are based on niobium and its alloys, particularly NbTi, $Nb_3Sn$ and $(NbTa)_3Sn$ are more often applied. Mostly Nb-based materials are used as wires to produce filamentary composite superconducting cables for grid network, wires and cables for high-power superconducting magnets, Superconducting Magnetic Energy Storage (SMES), compact generators, high-power compact electric motors, transformers, limiters, in electronic instrument manufacturing, etc. [1]. Suffice to say that each time after manufacturing of superconducting cable it is necessary to control check its critical parameters for those can vary a lot. [2]. The present article describes a new, original and relatively cheap method (ARC plasma-cluster method) for production of film-type superconducting materials which are supposed to be well utilized for superconducting cables production. Studies of main superconducting features as well as some physical and chemical properties are presented for new materials in the form of a coatings obtained by this method. The general target of the work is to introduce the flexibility of (ARC) plasma-cluster method for obtaining novel type of superconducting Nb materials so that it makes possible to manage critical parameters, simplify manufacturing process and at the same time reduce the costs of manufacturing superconducting materials for various applications.

*1) Description of coating formation.*

The original technology developed for deposition of superconductor coatings and film materials is based on well-known method of plasma flux generation from cathode spot (ARC-PVD). [3]. This method was suggested and implemented by Kraftonweg Ltd. Cathode spots are initiated on surface of operational block (cathode) do generate a plasma flux, which consists of plasma, vapor, and droplet phases. In result, plasma stream deposits onto surface of substrate, thus forming coating or film material. The process of NbTi layered coating deposition onto copper tubular conductor is schematically presented on Fig. 1.

The clusters (atomic quasi-crystals formed in the liquid micro-drop phase) containing molten droplet in sizes of 0.5 - 2.5μm serve as a basic to form the coating material. Contribution of cluster phase in overall mass transfer of evaporated material to substrate surface reaches 80% in value. The velocity of such clusters within plasma is about $10^2 \div 10^4$ cm/s. [4].

During coating process the left-hand extended evaporation system (see Fig. 1) generates Nb plasma flux, and right-hand produces the flux in Ti, respectively. Copper or steel tube serving base conductor for superconducting coating, performs both rotational and translational motion, as if screwed into vacuum chamber from below and coming out at the top. While the part of conductor is inside the vacuum chamber several layers of NbTi composition are formed at the top of the conductor. The rate of coating deposition for one cathode is in the range of $1 \div 3$μm/min, the temperature of deposition process is 700 - 800 ºC.

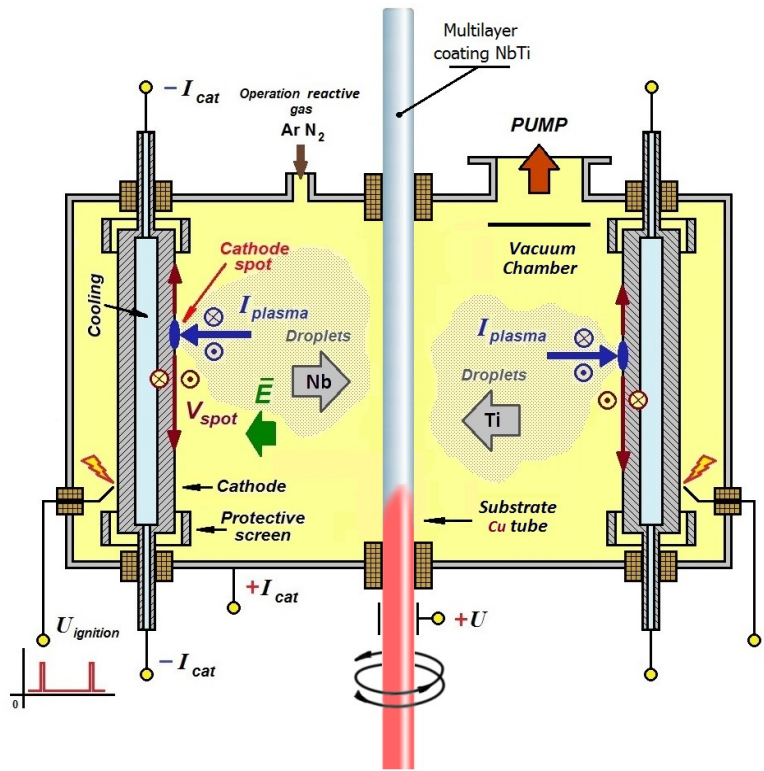

Fig. 1. Schematic diagram illustrating Ion-plasma-clusters method.

The spot clusters of molten metal created by cathode are moving inside plasma formation, and collide with electrons and ions, which charge clusters to the ambipolar potential. The value of cluster potential can be determined from the condition that electron and ion currents are equal on the surface of clusters moving in plasma cloud. Assuming that a nonisothermal plasma is formed in arcing stationary discharge at low pressure and plasma density is $n_{pl} = 10^8 \div 10^{12}$ cm$^{-3}$, electrons temperature is $T_e \approx 10$ eV, ions temperature is $T_i \leq 1$eV, to determine value of ions current per cluster, we can use Bohm formula:

$$j_i = 0{,}6 \times en_{pl}\sqrt{T_e/m_i}$$

In this case the value of cluster potential can be expressed as:

$$\varphi = (T_e/|e|)\ln(0{,}6\sqrt{2m_e/m_i}) \approx 52\text{V}$$

where $m_e$ and $m_i$ – mass of electron and ion, respectively; $e$ – electron charge.

Plasma ions, accelerating in the Debye layer, bombard surface of cluster with an energy corresponding to the potential value, and electrons are decelerated at cluster surface and only high-energy electrons from the tail of distribution function do reach the surface of cluster. The amount of energy transferred by electron and ionic components can be obtained from following formulas:

$$dE_e = n_{pl}\sqrt{\frac{kT_e}{2\pi m_e}}\, e^{-e\varphi/kT_e}\, kT_e\, S_c\, dt \qquad dE_i = 0{,}6 n_{pl}\sqrt{\frac{2kT_e}{m_i}}\, \varepsilon_i\, S_d\, dt$$

where $S_c$ – cluster surface area, $S_d = 4\pi(D_d/2 + D/2)^2$ - surface square of ion concentrating; $D_d\, 2 = R_d$ - thickness of Debye layer; $D$ – cluster size, $\varepsilon_i = |e\varphi|$ - energy of ions. Energy that electrons and ions give to surface of cluster amounts to $dE = dE_e + dE_i$.

Ions power is ten times greater than electrons power transferring to the cluster. Bombardment by electrons and ions of cluster surface leads to cluster heating. Simultaneously with transfer of energy to cluster its losses also occur due to infrared radiation and heat exchange with surroundings. Radiation cooling is described by the Stefan-Boltzmann law, which in for case can be expressed in the following form:

$$dE_r = a\sigma T^4 S_c\, dt$$

where $a$ - integral, average emissivity of cluster; $\sigma$ – Stefan-Boltzmann constant.

Heat exchange with surrounding space (neutral atoms) can be described by Newton's law which in our case has the form of:

$$dE_n = k n_n V_{Tn} S_c (T_n - T_c)\, dt$$

where $k$ – Boltzmann constant; $T_c$ – cluster temperature; $T_n$ – temperature of neutral atoms; $n_n$ and $V_{Tn}$ - density and thermal velocity of neutral atoms, respectively.

If temperature of single cluster changes by $-dT$, then energy of cluster will change by amount:

$$dEr + dEn = mcdT$$

where $c$ – heat capacity of cluster; $m$ – mass of cluster.

Applying heat balance equations for cluster, one can obtain a first-order differential equation and estimate cluster temperature $T$ [5].

Therefore, temperature of clusters moving in plasma can be modified by varying of plasma density $n_{pl}$. On the other hand the density of the plasma can be adjusted by applying external magnetic field to arc discharge gap.

2) *Structures and mechanical properties of* Nb, NbTi.

Cathodes of evaporation systems were made from the following metals: Niobium (Nb brand ASTM B392 purity 99,8%) and titanium (Ti brand DIN 3,7034 purity 99,6%). Three different types of film samples of Nb and NbTi coatings have been produced for testing.

Sample Nb - is coating film has been deposited on a copper substrate, foil.

Sample NbTi(V) - film, deposited from two combined plasma streams, was formed on copper and represents a bulk solid solution of Ti in Nb. Amount of Ti in that sample is 10%.

Sample NbTi(L) - layered film coating (deposited) with alternating Nb and Ti layers on a copper basis with subsequent separation of coating from substrate.

After deposition the samples of superconductive coatings have not been subjected to any additional treatment. Scanning electron microscope (SEM) image of Nb sample surface is shown in Fig. 2a, where a crystal growth texture of coating is clearly visible. Directed crystallites of medium cross-sectional size 5-10μm are combined into blocks, formed Dewsberry relief [6]. Cross-section of Nb coating (SEM image) is shown in Fig. 2b. Submicron crystalline structure is homogeneous, having an insignificant number of bulk point defects.

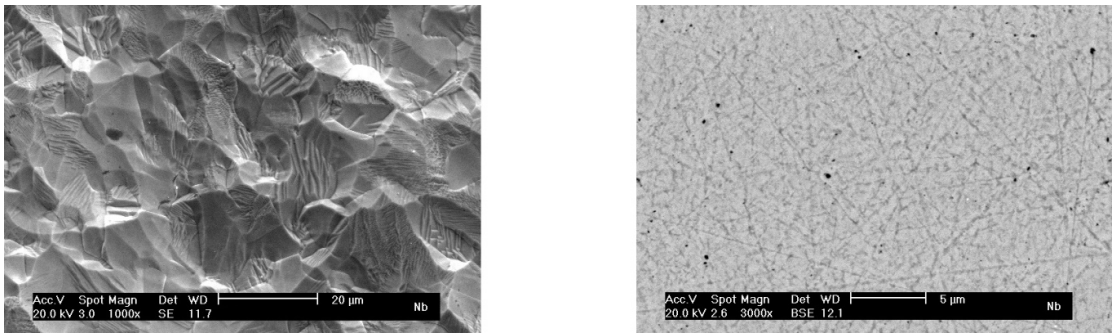

(a)            (b)

Fig. 2. Electron microscope images of surface (a) and cross-section (b) of Nb sample.

SEM images of surface and cross section of NbTi(V) sample visually indistinguishable from the Nb sample, by this reason we do not represent them here. SEM image of surface of NbTi(L) multilayer film sample is presented in Fig.3a. Relief and surface pattern of NbTi(L) sample is significantly different from Nb sample. Block structure of surface (average block size is 20 - 30 μm) has a homogeneous wavy structure with a wave period of 1 - 2 μm. Cross section of NbTi(L) sample (Fig. 3b.) has a clearly developed layered structure where the thicknesses of Nb layers substantially prevail the thicknesses of Ti layers.

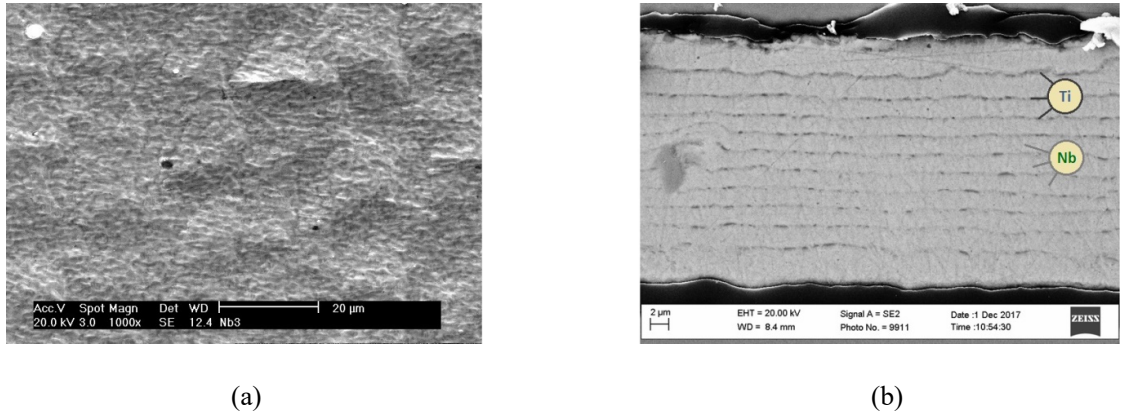

(a)                                                    (b)

Fig. 3. Electron microscope images of surface (a) and cross-section (b) of NbTi(L) sample.

XRD studies revealed that the direction of crystals growth is orthogonal to plasma-cluster flow direction. This gives unique possibility to control the structure of coating by changing angle between direction of plasma-cluster flow and substrate surface. To qualitatively evaluate the mechanical properties of coatings, we have subjected the Nb coatings to tensile stress testing. Mo sheet in 300 μm thickness was chosen as substrate material for coatings.

Tests were carried out on Tinius Olsen (H50KS) facility. The test results are presented below in (Table 1.)

Table 1. Parameters and mechanical properties of the samples used for mechanical tensile tests.

| Metal - base, Coating | Thickness, mm | Width, mm | Cross-section area, mm$^2$ | Length, mm | | Load, kN | Fracture strength, MPa | Relative extension, % | Relative necking, % |
|---|---|---|---|---|---|---|---|---|---|
| | | | | initial | final | | | | |
| Mo | 0,31 | 24,2 | 7,5 | 92,9 | 92,9 | 6,59 | 879 | - | - |
| Mo | 0,31 | 25,5 | 7,91 | 112,3 | 112,3 | 7,91 | 905 | - | - |
| Nb (film) | 0,17 | 9,0 | 1,53 | 25 | 26 | 0,148 | 96,7 | 4,0 | - |
| Nb (Mo) | 0,39 | 27,2 | 10,61 | 94,4 | 97,8 | 5,09 | 480 | 3,6 | 4,4 |
| NbTi(V) at Mo (*) | 0,35 | 20,9 | 7,32 | 94,2 | 94,7 | 6,32 | 863 | 0,5 | - |
| NbTi(V) at Mo (**) | 0,35 | 25,9 | 9,07 | 94,2 | 95,5 | 6,63 | 732 | 1,38 | 2,0 |

The following samples were used for mechanical tests:
1. Mo – fragment of sheet piece, not coated in rolled molybdenum M99,95-МП.
2. Nb(film) – sample coated with Nb film deposited on Cu foil, coating thickness ~ 80 μm. Surface hardness of Nb film $HV_{0.05} = 55 \div 67,2$ units.

3. Nb(Mo) – sample of Nb coating film deposited on Mo plate, coating thickness ~ 80 μm.

4. NbTi(V)Mo(*) – coating on molybdenum plate applied to one side, coating thickness ~ 40 μm.
5. NbTi(V)Mo (**) – coating at molybdenum plate applied to one side, coating thickness ~ 40 μm, with repeated loading.

This studies reveal rapture of both coatings and films. As expected, for Mo sample a brittle material rapture has been identified, graph showing dependence of stress on applied load is presented on Fig. 4.

Properties of Nb(film) on copper foil displayed in (Table 1) correspond to ones of cathode material. Form of tension diagram Fig. 5 and tests results for samples with thicknesses of about 1:1 ratio to the main substrate indicate a greater contribution of film to the properties of entire sample.

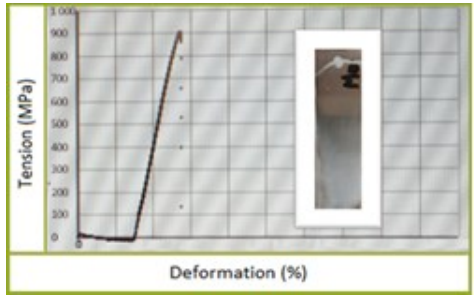
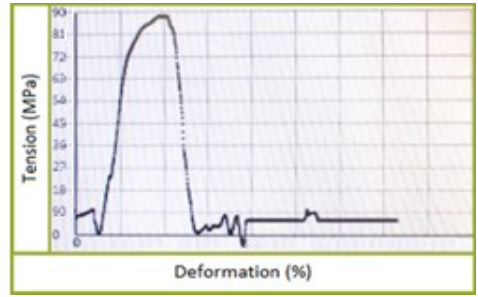

Fig. 4. Tension diagram for Mo sample.   Fig. 5. Tension diagram for Nb(film) sample.

Properties of Mo sample coated with Nb layer (Nb(Mo)) layer differ in strength. Strength of this sample is 1.9 times less(Fig. 6) which indicates the influence of the applied surface coating layer of Nb(Mo). Unlike uncoated Mo sample, the nature of rapture has changed – from brittle to brittle-viscous. Appearance of ductility indicates relative coherence crystal lattices of basic metal and film, which led to increased adhesion of coatings to substrate. Strength properties reduction of composite structure Nb(Mo) evidence the impact of Nb – films as a whole as well as good cohesive properties of this coating.

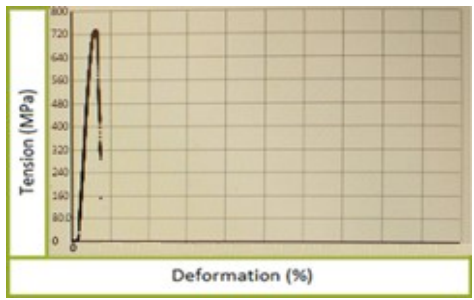
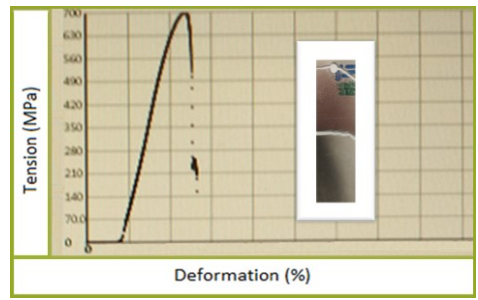

Fig. 6. Tension diagram for Nb(Mo).   Fig.7. Tension diagram for NbTi(V) Mo.

Mechanical properties of composition NbTi(V) Mo practically are not much different than strength of non-coated Mo sample (the difference is only 1.05 ... 1.2 times), which indicates an insignificant effect of deposited surface layer Fig.7.

The main brief conclusions drawn up from these results are presented below.
In the case of formation of relatively thick films, regardless the composition of substrate metal, the final structure practically has properties of evaporated metal.
Single layered and multilayered structures differ significantly in terms of texture and mechanical properties; controlling the structure (texture) one can affect mechanical properties of the coatings/films.
Coherence of film crystal lattices and base metal as well as ratio between adhesion-cohesive properties are the main factors allowing to determine growing structure through carrying out specific technology parameters of coating deposition.

*3) Cryogenic and electrical properties.*

Critical parameters of superconducting materials are key characteristics for their characterization and practical applications. These parameters can be obtained from electrical transport measurements and from magnetic measurements. Transport measurements usually applied for thin film samples with low critical currents, but in the case of the bulk samples or thick films with high critical current density transport measurements have high uncertainty and low reproducibility due to difficulties with fabrication of low resistive contacts and rather high electrical current through the sample during measurements. In current work we used magnetization measurements to derive main superconducting characteristics of our samples. Critical current and its magnetic field dependences were obtained from the magnetization hysteretic loops *M(B)*. Measurement of hysteretic magnetic field dependence of magnetization is often used for deriving critical current density $J_c$ and its dependence on magnetic field *B* in superconducting samples. Our calculations of the critical current density $J_c$ are based on frequently used Bean's model for long rectangular slab in parallel magnetic field [7, 7a, 7b,]:

$J_c(B) = 2 \Delta M(B) / (b (1 - b/3a))$,

where *ΔM* is a vertical width of the magnetization loop, *b* – slab thickness and *a* – slab dimension in perpendicular to magnetic field direction (*a* >> *b*).

Magnetic properties of superconducting coatings were evaluated using Quantum Design Magnetic Property Measurement System (MPMS XL7) based on superconducting quantum interference device. Measurements were carried out in parallel and perpendicular direction of magnetic field with respect to surface and coating. MPMS magnetometer was used with Reciprocating Sample Option providing sensitivity of approximately $5*10^{-9}$ EMU. Critical parameters of the superconducting samples were derived based on the measurement results. [8]..

MPMS XL7 magnetometer with External Device Control option was used for measurements of temperature dependence of electrical resistance. Keithley SourceMeter (model 2400) and Nanovoltmeter (model 2182A) where used for transport measurements.

Critical superconducting properties of obtained coatings (Nb, NbTi(V), and NbTi(L)) have been investigated in the temperature range 2 – 14K.

Nb coating films of different thicknesses have been evaporated on copper, stainless steel and silicon substrates. All films demonstrate similar characteristics with small deviation depending on the substrate material and film thickness. Temperature dependence of electrical resistance and critical current density at zero magnetic field for thin Nb coating (film Nb) on silicon substrate are are presented in Fig. 8a.

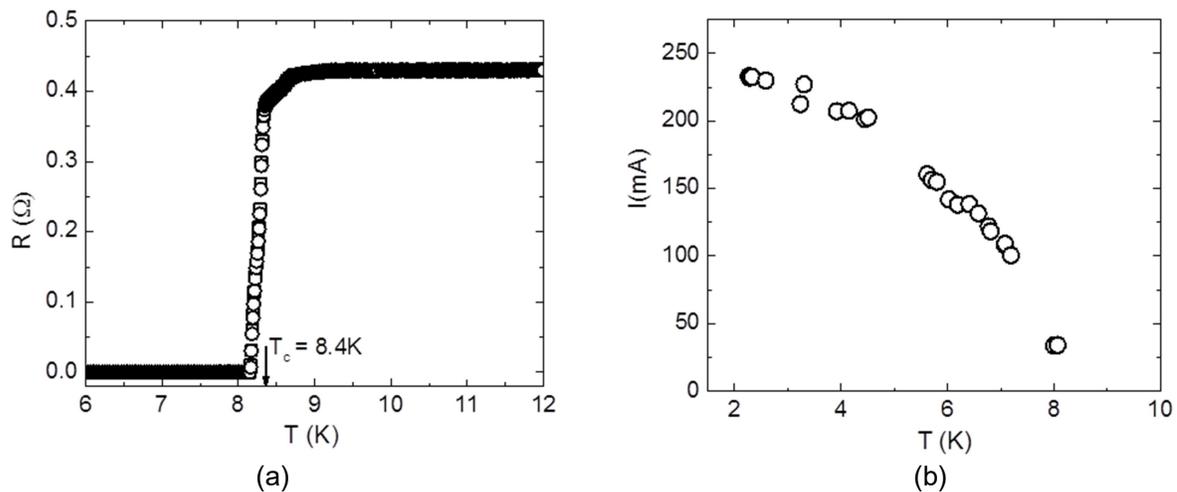

Fig. 8. Temperature dependence of electrical resistance (a) and transport critical current (b) for thin Nb(film) evaporated on silicon substrate.

In the case of the thick layers evaporated on copper substrate obtained Nb coating can be easily separated from substrate. Transition temperature for the thick films was derived from magnetic transition in low magnetic field (typically 25 Oe) and for few samples checked by transport measurements. Critical current density was derived from magnetic moment measurements using procedure described above.

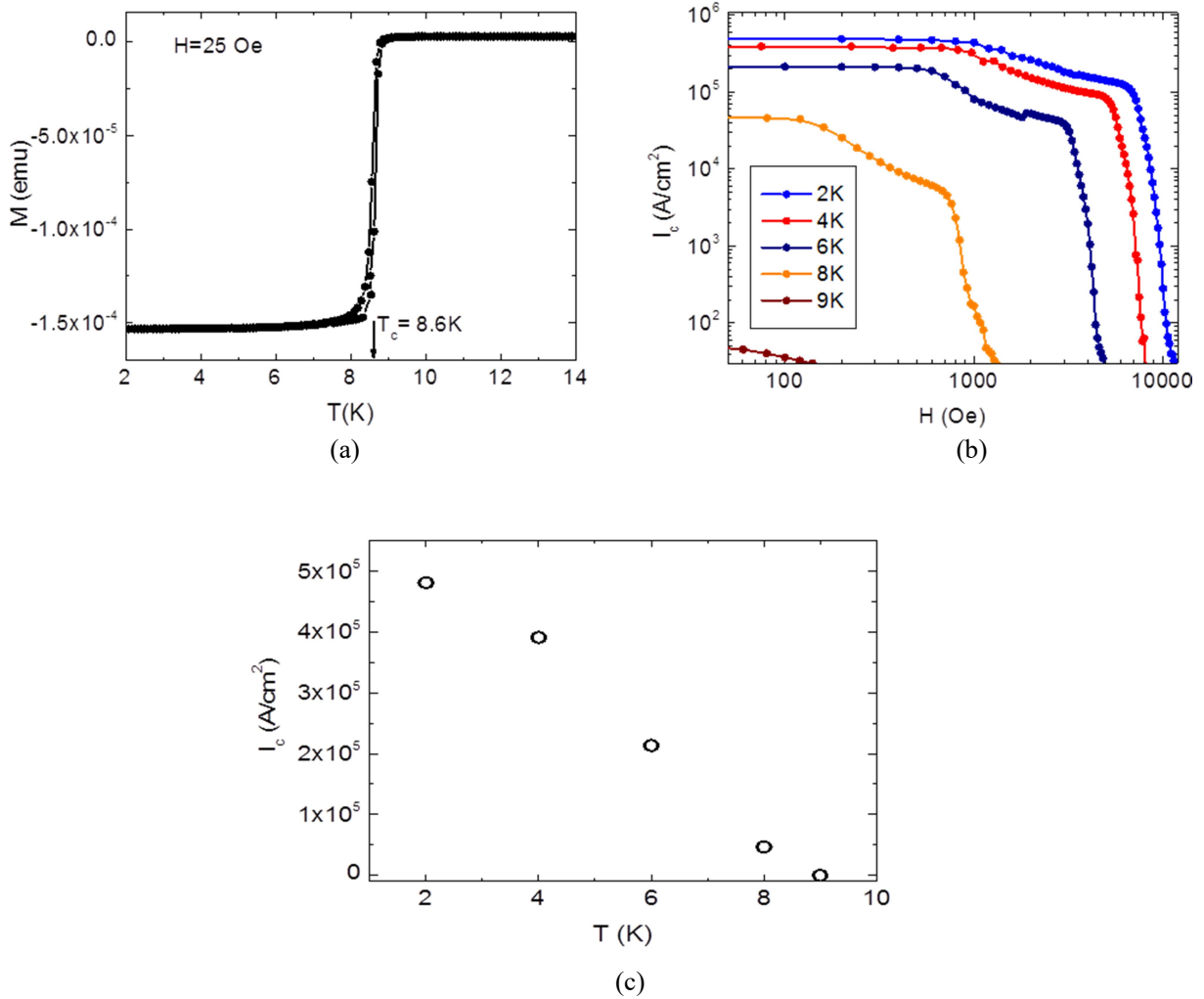

Fig. 9. Temperature dependence of the magnetic moment (a), magnetic field dependence of critical current density (b) and temperature dependence of critical current density at zero magnetic field (c) for thick Nb film.

Fig. 9 illustrates superconductor-normal metal transition and critical parameters for thick Nb sample evaporated on copper substrate. There is slight difference of the critical temperature and form of the transition curves for transport and magnetic measurements. This difference probably originate from significant difference of the film thicknesses and non-uniform thickness of the thin narrow film evaporated through mechanical mask. Value $T_c$ for clean Nb, manufactured by traditional industrial method is in range of $7 \div 9,2K$ [9], such a spread of values is associated primarily with crystallization process and following machining of superconducting material, products. Mechanical shaping (broaching) has a strong influence on crystal structure, in particular on surface crystal structure of superconductor. Value $J_c$ also depends quite strongly on way material was prepared [10]. Spread of critical current density values, depending on size of crystallites, may be in the range of $1\times10^3 \div 10^5$ A/cm2. Under such conditions, it is very difficult to fabricate superconducting materials with the same critical characteristics. Using ion plasma-cluster method crystal structure (and superconductor characteristics) can be reproduced with high accuracy. According with generally accepted microscopic theory of superconductivity, Bardeen–Cooper–Schrieffer theory (BCS), formation of electron pairs occurs when temperature of substance becomes less than a certain value (Tc), individual for each specific material with certain crystal structure. In our case introducing of Ti as impurity or additional layer leads to modification both local chemical composition and crystalline structure. Dependencies of critical values for coating NbTi(V) deposited under conditions of pooling, two plasma-cluster flows into one, and subsequent formation of condensate is shown in Fig. 10. Here we see a slight increase of Tc value (from 8.4-8.6K in the case of Nb film to 9.23K for NbTi(V)), which probably indicates change of chemical composition of the sample or influence of titanium atoms as local impurities on the characteristics of niobium. Observed critical temperature value of NbTi(V) sample corresponds to the maximum critical temperature of pure Nb.

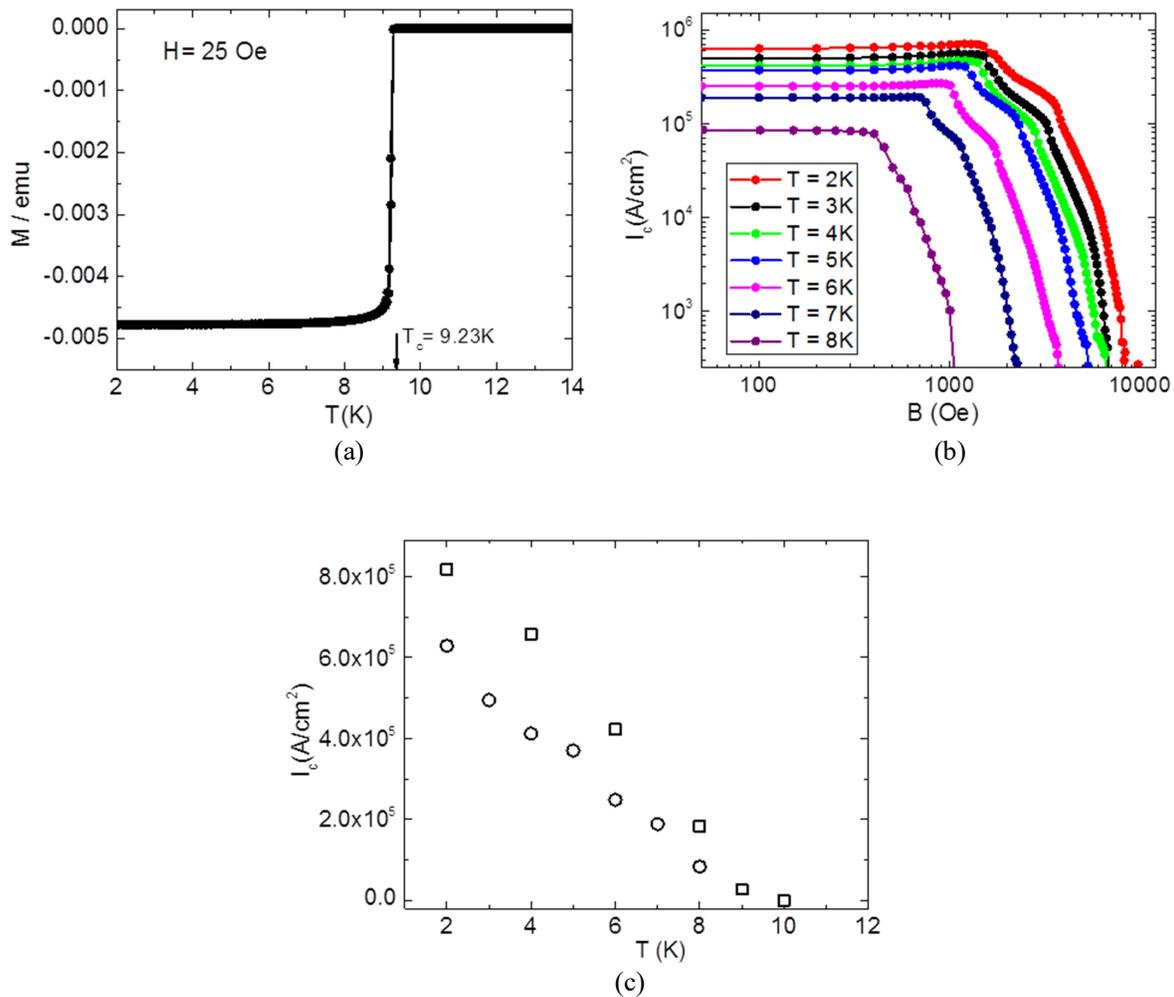

Fig. 10. Temperature dependence of magnetization (a) critical current dependence on magnetic field (b) and temperature (c) for NbTi(V) samples. Squares and circles corresponds to the different samples with different orientation of substrate during evaporation process.

Layered film coating with alternating Nb and Ti layers is characterized with similar critical temperature as NbTi(V) coating (see Fig. 11 a). However, there is a drastic difference in critical current density at zero magnetic field and its magnetic field dependence. Layered samples demonstrate both higher critical current density and weaker decay of critical current with increase of magnetic field.

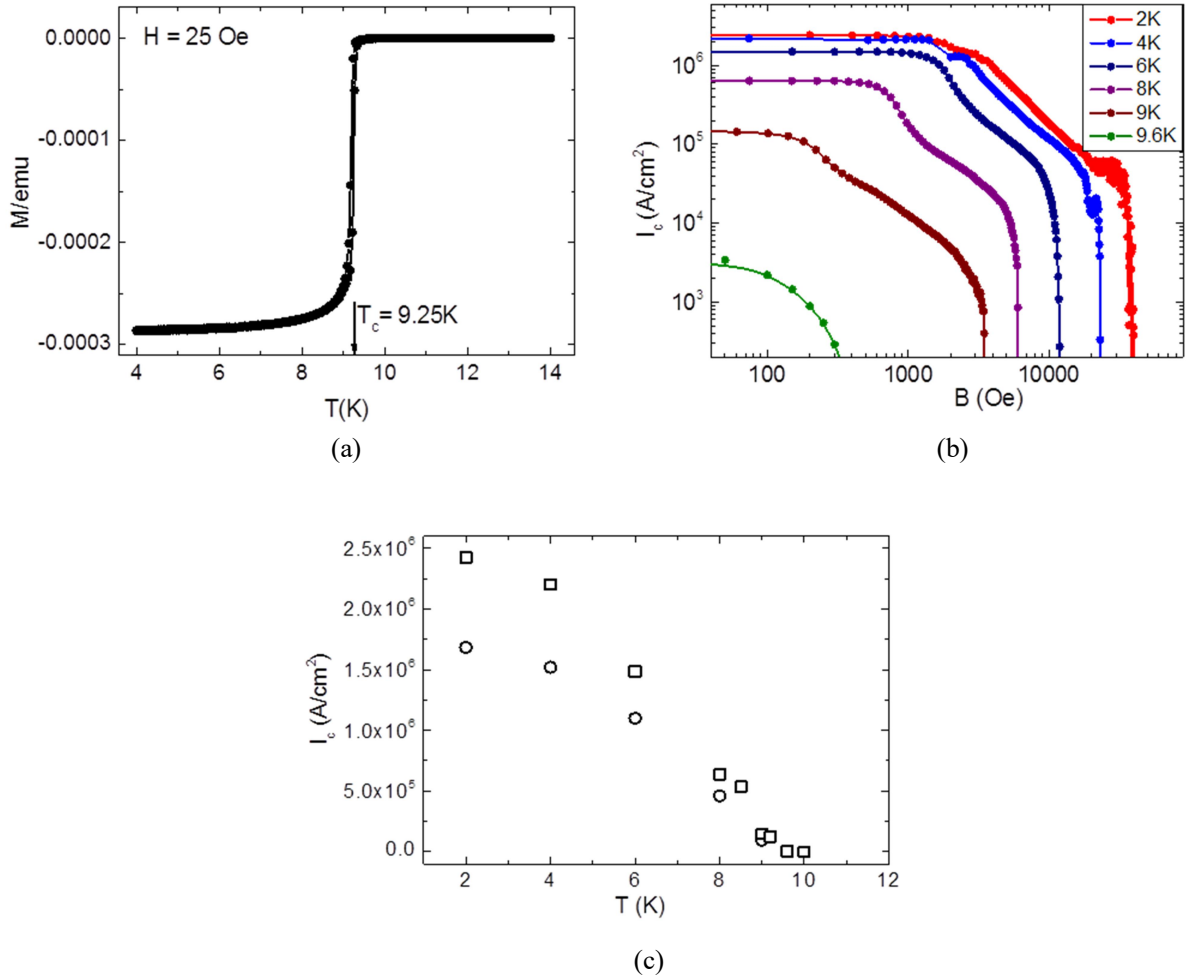

(a)

(b)

(c)

Fig. 11. Temperature dependence of magnetization (a) critical current dependence on magnetic field (b) and temperature (c) for NbTi(L) samples. Squares and circles corresponds to the samples evaporated on copper and stainless steel substrates.

It should be underlined that quantity of Ti in both coatings (NbTi(L) and NbTi(V)) is the same (about 10%), but form of Ti distribution in volume of Nb is quite different. In the case of NbTi(L) Ti and Nb the periodic layered structure perpendicular to the growths direction is well defined (see Fig. 3b). This structure is characterized by rather high critical current Jc = (1.5-2.5)×$10^6$ A/cm$^2$ in a magnetic field below 0.2 T. [10]

Superconductivity can be destroyed by electrical current flowing through superconductor. Such process of destruction superconductivity by critical current occurs in the form of avalanche. In transition from surface layer to normal state superfluid current, current near the surface of superconductor will begin to flow along the "core" of conductor. Reducing effective radius of superconductor will result at its surface will become even larger H∝1/r, which leads to loss of superconductivity and screening of the magnetic field preventing its penetration into interior.

For gap Δ=10K critical current density will be Jc ≈ 2×$10^8$ A/см$^2$, this value is maximum possible within the framework of (BCS).

Superconducting current flows in a layer of thickness equal $\lambda \approx 1000$Å $= 10^{-5}$ см, or 0,1 µm, which determines minimum possible thickness of superconducting layer of conductor. Based on magnetic flux quantization condition, and conditions for simultaneous interaction of vortices in volume of a superconductor we take minimum possible distance between the vortices-fluxons at least 3500 Å. This condition must be taken into account when determining minimum thickness of Nb layer in coatings.

As a result, the dependences of the critical values obtained for our coatings are in agreement with well-known mathematical expressions describing superconductive properties.

We have demonstrated that introduction of Ti into Nb matrix in various forms does affect superconductive properties of final composite structures. The largest enhancement of superconducting parameters was observed on the layered Nb-Ti system. The choice of optimal thickness and shape of Ti inclusions remains undefined and requires additional studies in this direction. So far we assume that density of superconducting current relates the value of magnetization of layers which are in normal state. Magnetization, in turn, is related to the actual structure (texture) of surface layers, and in first approximation, can be estimated using the Weiss scale [11].

*4) Conclusions.*

1) Original, well-controlled, technology for obtaining superconductive coatings and/or films basing on low-pressure (ARC) discharge in metal vapors is presented.

2) Coatings and films obtained by above method feature both high manufacturability and high reproducibility. Coatings can be put on intricate geometric shapes. Coatings require no further additional processing before installation and usage in final product. Such materials can be bent at an angle 180° without any damage and loss of superconductive properties, subjected to molding or reeling at practically any diameter.

3) A very high value of critical current density ($J_c$) was found for NbTi(L) coating. Reduction of Slope reduction of dependence curve ($J_c$) as function of external of magnetic field ($B_c$) have been found.

4) Simple, efficient, well-operated, and relatively cheap method for producing superconductive niobium-based coatings could be proposed as new basic method for producing of superconducting metallic materials.

*Acknowledgments*

*The work has been finically supported by Kraftonweg Oy.*

*Authors appreciate the help of staff of Ota Nano, Low Temperature Laboratory of Aalto University in Critical Temperature testing.*

*Our special thanks to professional engineer-metallurgist Mr. A. Zhivushkin for the help and advising provided.*

*References:*

[1]  C. Li, D. C. Larbalastier: Development of high critical current densities in niobium 46.5 wt% titanium, Cryogenics 27, 171 (1987)

[2]  Niobium-Titanium Superconducting Wires: Nanostructures by Extrusion and Wire Drawing
By Peter J. Lee and David C. Larbalestier   The Applied Superconductivity Center, The University of Wisconsin-Madison USA.

[3]  Investigations of vacuum ARC cathode spots with high temporal and spatial resolution. B. Bochkarev and Aidar M. Murzakaev Institute of Electrophysics , Russian Academy of Sciences
620046 , Komsomol'skaya St. 34 , Ekatherinburg , Russia

[4]  McClure G.W. Plasma expansion as a cause of metal displacement in vacuum arc spots // J. Appl. Phys. -1974. –Vol. 45. - №5. – P. 2078-2084

[5]  Dynamics of drop phase in plasma of  low pressure ARC discharge.
A.A. Bizyukov, E.V. Romashchenko*, K.N. Sereda, A.D. Chibisov, A.Y. Kashaba
Kharkiv National University, 31 Kurchatov Ave., Kharkiv, 61108, Ukraine;

[6]  Kelly, A.; Groves, G. W.   Crystallography and Crystal Defects.  p.496
Published by Longman, London (1970)

[7]  Bean, C. P. Magnetization of high-field superconductors. Rev. Mod. Phys. 36, 31–39

[7a]  D.-X. Chen, R.B. Goldfarb, Kim model for magnetization of type-II superconductors, J. Appl. Phys. **66** (1989) 2489.

[7b]  T.H. Johansen, H. Bratsberg, Critical-state magnetization of type-II superconductors in rectangular slab and cylinder geometry, J. Appl. Phys. **77** (1995) 3945.

[8]   An Extended Critical State Model: Asymmetric Magnetization Loops and Field Dependence of the Critical Current of Superconductors D. M. Gokhfeld  Kirensky Institute of Physics, Siberian Branch of the Russian Academy of Sciences, Akademgorodok 50–38, Krasnoyarsk, 660036 Russia

[9]    O. Henkel, E. M. Sawitzkij (Eds): Supraleitende Werkstoff   VEB Deutscher (Verlag fur Crundstoffindustrie, Leipzing 1982 ) (in German)

[10]  Vortex Pinning in Poly-and Monocrystalline, Superconductors of Niobium.
Article in The Physics of  Metals and Metallography · January 1982
V. G. Glebovsky ,  L. Ya. Vinnikov  Institute  of Solid State         …

[11]  R. J. Weiss   Solid State Physics for Metallurgists. PP Oxford-London-New York-Paris, 1963.